\documentstyle[epsfig]{aprim}

\newif\ifAMStwofonts

\ifoldfss
  \ifCUPmtlplainloaded \else
    \NewTextAlphabet{textbfit} {cmbxti10} {}
    \NewTextAlphabet{textbfss} {cmssbx10} {}
    \NewMathAlphabet{mathbfit} {cmbxti10} {} 
    \NewMathAlphabet{mathbfss} {cmssbx10} {} 
  \fi
  \ifAMStwofonts
    \ifCUPmtlplainloaded \else
      \NewSymbolFont{upmath} {eurm10}
      \NewSymbolFont{AMSa} {msam10}
      \NewMathSymbol{\upi}     {0}{upmath}{19}
      \NewMathSymbol{\umu}     {0}{upmath}{16}
      \NewMathSymbol{\upartial}{0}{upmath}{40}
      \NewMathSymbol{\leqslant}{3}{AMSa}{36}
      \NewMathSymbol{\geqslant}{3}{AMSa}{3E}

    \fi
  \fi
\fi 

\ifnfssone
  \newmathalphabet{\mathit}
  \addtoversion{normal}{\mathit}{cmr}{m}{it}
  \addtoversion{bold}{\mathit}{cmr}{bx}{it}
  \newmathalphabet{\mathbfit} 
  \addtoversion{normal}{\mathbfit}{cmr}{bx}{it}
  \addtoversion{bold}{\mathbfit}{cmr}{bx}{it}
  \newmathalphabet{\mathbfss} 
  \addtoversion{normal}{\mathbfss}{cmss}{bx}{n}
  \addtoversion{bold}{\mathbfss}{cmss}{bx}{n}
  \ifAMStwofonts
    \ifCUPmtlplainloaded \else
      \UseAMStwoboldmath
      \makeatletter
      \new@mathgroup\upmath@group
      \define@mathgroup\mv@normal\upmath@group{eur}{m}{n}
      \define@mathgroup\mv@bold\upmath@group{eur}{b}{n}
      \edef\UPM{\hexnumber\upmath@group}
      \new@mathgroup\amsa@group
      \define@mathgroup\mv@normal\amsa@group{msa}{m}{n}
      \define@mathgroup\mv@bold\amsa@group{msa}{m}{n}
      \edef\AMSa{\hexnumber\amsa@group}
      \makeatother
      \mathchardef\upi="0\UPM19
      \mathchardef\umu="0\UPM16
      \mathchardef\upartial="0\UPM40
      \mathchardef\leqslant="3\AMSa36
      \mathchardef\geqslant="3\AMSa3E
    \fi
  \fi
\fi 

\ifnfsstwo
  \DeclareMathAlphabet{\mathbfit}{OT1}{cmr}{bx}{it}
  \SetMathAlphabet\mathbfit{bold}{OT1}{cmr}{bx}{it}
  \DeclareMathAlphabet{\mathbfss}{OT1}{cmss}{bx}{n}
  \SetMathAlphabet\mathbfss{bold}{OT1}{cmss}{bx}{n}
  \ifAMStwofonts
    \ifCUPmtlplainloaded \else
      \DeclareSymbolFont{UPM}{U}{eur}{m}{n}
      \SetSymbolFont{UPM}{bold}{U}{eur}{b}{n}
      \DeclareSymbolFont{AMSa}{U}{msa}{m}{n}
      \DeclareMathSymbol{\upi}{0}{UPM}{"19}
      \DeclareMathSymbol{\umu}{0}{UPM}{"16}
      \DeclareMathSymbol{\upartial}{0}{UPM}{"40}
      \DeclareMathSymbol{\leqslant}{3}{AMSa}{"36}
      \DeclareMathSymbol{\geqslant}{3}{AMSa}{"3E}
    \fi
  \fi
\fi 

\ifCUPmtlplainloaded \else
  \ifAMStwofonts \else 
    \def\upi{\pi}
    \def\umu{\mu}
    \def\upartial{\partial}
  \fi
\fi

\title[Taniguchi, Y., et al.]{Early Stage of Galaxy Formation}

\author[Taniguchi et al.]
       {Y. Taniguchi$^1$, T. Nagao$^{2, 3}$, M. Ajiki$^1$,
        Y. Shioya$^1$, S. S. Sasaki$^1$, \& T. Murayama$^1$  \\
        $^1$Astronomical Institute, Graduate School of Science, Tohoku University,
        Aramaki, Aoba, Sendai 980-8578, Japan \\
        $^2$National Astronomical Observatory of Japan, 2-21-1 Osawa, Mitaka,
        Tokyo 181-8588, Japan \\
        $^3$INAF --- Osservatorio Astrofisico di Arcetri, Largo Enrico Fermi 5, 50125 
         Firenze, Italy}
\date{}

\pagerange{\pageref{firstpage}--\pageref{lastpage}}
\pubyear{2005}

\begin{document}

\maketitle

\label{firstpage}

\begin{abstract}
We discuss on the early stage of galaxy formation
based on recent deep surveys for very high-redshift galaxies,
mostly beyond redshift of 6. These galaxies are observed to be
strong Lyman$\alpha$ emitters, indicating bursts of massive star
formation in them. The fraction of such star-forming system
appears to increase with increasing redshift. On the other hand, 
the star formation rate density derived from Lyman$\alpha$ emitters
tends to decrease with increasing
redshift. It is thus suggested that the major epoch of initial
starbursts may occur around $z \sim$ 6 -- 7. In order to understand
the early stage of galaxy formation, new surveys for galaxies
beyond redshift of 7 will be important in near future.
\end{abstract}

\begin{keywords}
galaxies: formation --- galaxies: evolution
\end{keywords}


\section{Introduction}

The formation and evolution of galaxies have been intensively
studied from both observational and theoretical points of view
for these two decades. The progress in this research field 
can be attributed to several deep optical survey programs.
For example, the galaxy evolution from $z=1$ to the  present day 
has been studied based on CFRS (= Canada-France Redshift Survey:
Lilly et al. 1995), SDSS (= Sloan Digital Sky Survey: York et al. 2000),
2dF (= 2 Degree Field Survey: Colless et al. 2001), and so on.
These surveys are also useful in exploring the nature of large scale
structures.
On the other hand, the galaxy evolution from $z \sim 6$ to $z \sim 1$
has been studied based on deep high-resolution imaging surveys
with the Hubble Space Telescope such as HDF (= Hubble Deep Fields:
Williams et al. 1996, 2000), GEMS (= Galaxy Evolution from Morphology 
and SEDs: Rix et al. 2004), GOODS (= Great Observatories Origins
Deep Survey: Giavalisco et al. 2004a), HUDF (= Hubble Ultra Deep
Field: Beckwith et al. 2005; Thompson et al. 2005) and COSMOS 
(= Cosmic Evolution
Survey: Scoville et al. 2004, 2006; see also Taniguchi et al. 2005b). 
These huge high-quality data sets are highly useful in investigating
the evolution of galaxies from $z \sim 6$ to the present day.
Furthermore, 8-10m class ground-based telescopes also contribute
to the understanding of galaxy evolution thanks to their
great spectroscopic capability for large samples of very faint galaxies (e.g.,
Cowie et al. 1996; Steidel et al. 1996, 1999; Cohen et al. 1999 [Caltech Faint
Galaxy Survey]; Abraham et al. 2004 [GDDS = Gemini Deep Deep Survey];
Vogt et al. 2005 [DEEP Groth Strip Survey]; 
Wirth et al. 2004 [The Team Keck Treasury Redshift Survey];
Le F\`evre et al. 2005 [VIMOS VLT Deep Survey]).

These important investigations have made great contributions
to the understanding of dynamical, chemical, and luminosity evolution of
galaxies in various galaxy environs from high redshift to the present day. 
For example, the cosmic star-formation history over a period of
10 billion years in the universe has been investigated systematically 
(e.g., Madau et al. 1996; Steidel et al. 1999; Giavalisco et al. 2004b;
Dickinson et al. 2004). It is now also accepted that the nature of galaxy
evolution is understood as less massive galaxies tend to have longer
star formation timescale (i. e., the down sizing: e.g., Cowie et al. 1996;
Heavens et al. 2004; Kodama et al. 2004; Treu et al. 2005). 
It is also interesting to note that low-redshift galaxies are classified 
into two distinct families at a stellar mass of $3 \times 10^{10} M_\odot$
(Kauffmann et al. 2003), suggesting that the galaxy evolution is related
to the surface mass density.
Moreover, most optical surveys have been coordinated with surveys at 
other wavelengths, e.g., from mid- through far-infrared to submm, and
X-ray. These multiwavelnghth surveys have explored the dark side
of galaxy evolution (e.g., Hughes et al. 1998; Barger et al. 1998; 
Chapman et al. 2003; Elbaz et al. 1999),
and the co-evolution between the galactic spheroidal system    
and the central supermassive black hole (e.g., Barger et al. 2001).

One of remaining important issues related to galaxies is the formation of
galaxies. Since the epoch of first stars (i.e., Population III objects)
may be at $z \sim$ 10 -- 30 (e.g., Loeb \& Barkana 2001), probing very
high-redshift universe is absolutely necessary to understand the physical
process of galaxy formation.
Since the great success of the Hubble Deep Field project (Williams
et al. 1996), a large number of high-redshift galaxies have been
discovered so far to date (e.g., Taniguchi et al. 2003; Spinrad 2004).
In particular, deep optical imaging surveys with a narrowband (NB) filter
have been very successful in finding star-forming galaxies at
$z \sim$ 5 -- 6 (Rhoads et al. 2000; Malhotra \& Rhoads 2004;
Hu et al. 2002, 2004; Ajiki et al. 
2003; Ouchi et al. 2003, 2005; Shimasaku et al. 2004; Santos et al. 2004;
see for reviews, Taniguchi et al. 2003; Spinrad 2004). Their star formation
properties and the luminosity function have been intensively studied 
for these several years. Large samples of Lyman break galaxies at $z \sim 6$
are also investigated in detail (Giavalisco et al. 2004b; Dickinson et al.
2004; Stanway et al. 2004; Bouwens et al. 2004; Shimasaku et al. 2005). 
In this review, we give a summary of observations of galaxies beyond $z=6$
to investigate the early stage of galaxy formation.

\section{Galaxies beyond Redshift of 6}

Galaxies beyond $z =6$ can be found  by the following two methods; 
(1) searches for Ly$\alpha$ emission (e.g., Hu et al. 2002; Taniguchi et al. 2005a
and references therein), and (2) broad-band color selections
such as $I$-band dropouts (e.g., Giavalisco et al. 2004b; Dickinson et al. 2004;
Stanway et al. 2005; Bouwens et al. 2005a). 
Galaxies found by the first method are referred as Lyman$\alpha$ emitters (LAEs)
while those found by the second one are as Lyman break galaxies (LBGs).
However, both LAEs and LBGs at $z > 6$ could be the same population of forming 
galaxies even though their selection methods are different.
Indeed, some of LBGs selected as $I$ dropouts show strong Ly$\alpha$ 
emission line (e.g., Nagao et al. 2004, 2005b; Stanway et al. 2004).
In this section, we discuss the nature of galaxies at $z > 6$ whose redshifts
are confirmed by optical spectroscopy.

We give a list of 18 galaxies at $z > 6$ compiled from the literature 
by the end of July, 2005. Among the 18 objects, 13 objects have been
found with the narrowband imaging technique. Although survey redshift ranges
are limited because of strong OH airglow emission lines,
the narrowband method appears to be very efficient to select  LAEs;
note that a NB filter centered at $\lambda \approx$ 816 nm is used 
to find LAEs at $z \approx 5.7$ while that centered 
at $\lambda \approx$ 921 nm is used 
to find LAEs at $z \approx 6.6$ (e.g., Taniguchi et al. 2003). 

One interesting case is 0226$-$04LAE found by Cuby et al. (2003).
Although this object was selected as a NB-excess object,
its excess emission was found to be redshifted UV continuum.
Their optical spectroscopy identified this object as a LAE at $z = 6.17$.
Another interesting case is SEXSI-SER found by Stern et al. (2005).
This object, a LAE at $z=6.545$, is a serendipitously identified during the course
of their follow-up observations of optical counterpart of X-ray sources.
Such serendipitous identifications of high-$z$ galaxies are also reported in
Dey et al. (1998) and Dawson et al. (2002).

The broadband color selection method brought us four objects 
at $z>6$. All these objects show a strong Ly$\alpha$ emission line.
Among them, the three objects found by Nagao et al. (2004, 2005b) 
were identified as red $i^\prime - z^\prime$ objects and thus 
suspected to be LBGs at $z \sim 6$. However, their follow up 
spectroscopy identified them as strong LAEs at $z$ = 6.33, 6.04, 
and 6.03. This means that most of their $z^\prime$ fluxes come not from
their UV continuum but from Ly$\alpha$ emission.

The star formation rate (SFR) of the galaxies at $z >6$ given in Table 1
is several $M_\odot$ y$^{-1}$ on average (e.g., Taniguchi et al. 2005a).
The star formation rate density (SFRD) at $z \approx$ 6.6 inferred from 
nine LAEs studied by Taniguchi et al. (2005a) is estimated to be
$\sim 6 \times 10^{-4}$ $M_\odot$ y$^{-1}$ Mpc$^{-3}$. This is 
smaller by two orders of magnitude than those derived from LBGs
at $z \sim 6$ (e.g., Giavalisco et al. 2004b; Dickinson et al. 2004;
Stanway et al. 2004; Bouwens et al. 2004). Even if we correct for 
extinction for Ly$\alpha$ emission (i.e., a factor of a few) and 
an integrate with a certain Ly$\alpha$ luminosity function
(i.e., a factor of a few), the SFRD derived from the $z \approx 6.6$ 
sample is still smaller by one order of magnitude than that 
derived from the $z \sim 6$ LBG samples. The intergalactic medium (IGM)
cannot be re-ionized by the LAEs found to date; note that LBGs
found at $z \sim$ 6 also cannot reionize the IGM.

It is worth noting that the star formation rate density derived from 
LAEs tends to decrease with increasing redshift (e.g., Taniguchi et al.
2005; Yamada et al. 2005). This suggests that the major epoch of initial
starbursts may occur around $z \sim$ 6 -- 7.

\begin{table*}
\begin{center}
\begin{tabular}{clccc}
\hline \hline
No. & Name & $z$ & Method$^1$ & Ref.$^2$  \\
\hline
1 & SDF J132522.3+273520 & 6.597 & NB &  1 \\
2 & SDF J132432.4+271647 & 6.580 & NB & 1 \\
3 & SDF J132418.3+271455 & 6.578 & NB & 1 \\
4 & SDF J132518.8+273043 & 6.578 & NB & 1, 2 \\
5 & HCM-6A               & 6.56  & NB & 3 \\
6 & SDF J132408.3+271544 & 6.554 & NB & 1 \\
7 & SEXSI-SER            & 6.545 & SER & 4 \\
8 & SDF J132352.7+271622 & 6.542 & NB & 1 \\
9 & SDF J132415.7+273058 & 6.541 & NB & 1, 2 \\
10 & SDF J132353.0+271631 & 6.540 & NB & 1 \\
11 & LALA142442.2+353400 & 6.535 & NB & 5 \\
12 & KCS 1166            & 6.518 & Grism & 6 \\
13 & SDF J132418.4+273345 & 6.506 & NB & 1 \\
14 & SDF J132440.6+273607 & 6.330 & CS & 7 \\
15 & GOODS-N $i^\prime$-drop No. 6 & 6.24 (?) & CS & 8 \\
16 & 0226-04LAE         & 6.17   & NB & 9 \\
17 & SDF J132442.5+272423 & 6.04 & CS &10 \\
18 & SDF J132426.5+271600 & 6.03 & CS &10 \\
\hline \hline
\end{tabular}
\end{center}
\begin{center}
\caption{A list of galaxies beyond $z$=6. 
$^1$Method: NB = narrowband search, SER = serendipitous discovery,
Grism = Grism imaging spectroscopy, \& CS = 
broadband color selection. 
$^2$References: 1. Taniguchi et al.
2005, 2. Kodaira et al. 2003, 3. Hu et al. 2002, 4. Stern et al. 2005,
5. Rhoads et al. 2004, 6, Kurk et al. 2004, 7. Nagao et al. 2004, 8.
Stanway et al. 2004, 9. 
Cuby et al. 2003, and 10. Nagao et al. 2005b.}
\end{center}
\end{table*}

\section{Galaxies beyond Redshift of $\sim$ 7}

In this section, we summarize recent results on searches for objects 
beyond $z \sim 7$. The Lyman break and Ly$\alpha$ emission are
redshifted to 730 nm and 973 nm, respectively for an object at $z=7$.
In particular, it becomes difficult to use the Ly$\alpha$ emission as a probe of
such very high-$z$ galaxies because the sensitivity of CCD cameras is poor. 
Even if we use the Lyman break and continuum depression at wavelengths shorter
than 121.6 nm, such galaxies can be detected only in $z^\prime$ band
in the optical. Therefore, near and mid infrared data become much more important 
for investigations of such very high-$z$ galaxies.
Another observational difficulty should come from that such very high-$z$ galaxies 
are inevitably faint. Therefore, we need new techniques to investigate 
galaxies beyond $z \sim 7$ in principle.
 
One promissing method is searches for gravitationally amplified objects.
Gravitational lensing caused by a relatively nearby massive cluster of
galaxies is very useful in searching for faint, very high-$z$ galaxies
(e.g., Ellis et al. 2001; Hu et al. 2002; Santos et al. 2004).
For example, nine LAEs at $z \sim 6.6$
are found in a field including Abell 370 while no detection of such galaxies
in HDF (Cowie 2004).

The most probable galaxy at $z \sim 7$ is a triple-imaged galaxy found in 
a field of Abell 2218 ($z_{\rm cluster} = 0.1775$) (Kneib et al. 2004).
This source is also detected at 3.6 and 4.5 $\mu$m using the Spitzer Space Telescope
(Egami et al. 2005), indicating its photometric redshift of $z_{\rm phot}
\simeq$ 6.6 -- 6.8.
Its age is estimated to be 50 -- 450 Myr with the star formation rate of 
0.1 -- 5 $M_\odot$ y$^{-1}$. The estimated mass is $\sim 10^9 M_\odot$, being smaller 
than those of typical LBGs at $z \sim 3$. The finding such a small-mass (or, subgalactic)
system is indeed attributed to the large amplification factor of the lensing ($\sim 25$).

Other probable very high-$z$ galaxies have been found in the UDF. 
Bouwens et al. (2004b) found 5 probable objects at $z \sim$ 7 -- 8 among the
$z_{850}$ dropouts in the UDF (see also Yan \& Windhorst 2004). 
Then, Bouwens et al. (2005) also found 3 probable objects at $z \sim 10$
among the eight $J$ dropouts in UDF. All of them are too faint (e.g.,
$J \sim 27$ for $z_{850}$ dropouts and $H \sim 27$ for $J$ dropouts) to be
observed spectroscopically. 
The star formation rate density (SFRD)at $z \sim$ 7 -- 8 and $z \sim$ 10 is  
$\sim 10^{-2.5}$ and $\sim 10^{-3}$ $M_\odot$ y$^{-1}$ Mpc$^{-3}$ in the 
WMAP cosmology, respectively. These data suggest the continuous decline of  
SFRD from $z \sim 3$ to $z \sim 10$; i.e., SFRD($z \sim 10$)/SFRD($z \sim 3$)
$\simeq 0.2$ (Bouwens et al. 2005).

Any challenges with ground-based telescopes have been giving null results on 
searches for galaxies beyond $z \sim 7$. As noted above, most such very
high-$z$ objects could be too faint. Since strong OH airglow emission lines 
make it difficult to carry out deep imaging surveys at wavelengths longer than
700 nm.  In the case of the Subaru Deep Field (Kashikawa et al. 2004), 
although a few tens galaxies at $z \sim$  6 -- 6.6 have been found (Shioya et
al. 2005a; Shimasaku et al. 2005; Taniguchi et al. 2005), no object has been
found at $z > 6.7$ (Shioya et al. 2005b). This is mainly due to the
shallow survey depth at $z^\prime$ band; i.e., $z^\prime_{\rm lim} \simeq 26$.

In the near infrared (NIR) window, any ground-based observations share the
same problem because of strong airglow emission lines and strong thermal
background in NIR. 
Therefore, NIR broad-band imaging surveys may have difficulty in finding 
very high-$z$ galaxies at $z > 7$. One remaining technique is again the
narrowband imaging survey. Recently, Willis \& Courbin (2004) made such a
narrowband deep survey using a NB filter centered at $\lambda$ = 1.187 $\mu$m
using the VLT/ISAAC facility; i. e., the corresponding Ly$\alpha$ redshift is 
$z \sim 9$. Although their survey depth was down to $\sim 3 \times 10^{-18}$ 
erg s$^{-1}$ cm$^{-2}$, they  found no LAE.

Another search for galaxies at $z > 7$  was also made by P\'ello et al. (2004)
as their follow-up observations of gravitationally amplified objects
in a field of Abell 1835. Although they found a possible signature of
Ly$\alpha$ emission at $\lambda$ = 1.337 $\mu$m, indicating the redshift of
$z \simeq 10$, later investigators did not confirm this finding (Weatherley
et al. 2004; Bremer et al. 2004; see also Lehnert et al. 2004).
This also suggests the technical difficulty in ground-based observations of
such very high-$z$ galaxies. However, their challenges and discussion given in
the above papers will be useful for our future investigations
 (see also Cen, Haiman, \& Mesinger 2005).

\section{Future Searches for Objects beyond Redshift of 7
         and Population III Objects}

One of the most important issues remained unsettled is how galaxies 
were assembled in their early phase and what were first objects 
(i. e., Population III objects) 
after the dark age. These problems are also related to the understanding
physical processes of the cosmic reionization (e.g., Loeb \& Barkana 2001).

Recently, Nagao et al. (2005a) tried to detect intense He {\sc ii} $\lambda$1640
emission line in one of the brightest LAEs found in the SDF 
(SDF J132440.6+273607:  Nagao et al. 2004).
If the photoionization would be dominated by Population III stars in this LAE, 
they could detect He {\sc ii}, but failed. However, since a number of 
apparently bright LAEs beyond $z =6$ will be found in near future, this kind
of challenges will be very important to understand the nature of star formation 
properties in such very high-$z$ young galaxies. 

In future observations of very high-$z$ galaxies, HST will have advantage
with respect to any ground-based telescopes because of low noise in the NIR
window. However, in order to proceed investigations of very high-$z$ galaxies,
a new, wide-field NIR camera and spectrograph for HST is necessary.
If a very wide-field NIR camera is available on 8m class telescopes on the ground,
deep (mostly narrowband) imaging surveys will be able to find some very high-$z$
galaxies from a statistical point of view.  

NIR cameras currently available on 8m class telescopes are;
1) NIRI on Gemini (FOV = 2$^\prime \times 2^\prime$),
2) ISAAC on VLT (FOV = 2.5$^\prime \times 2.5^\prime$),
3) CISCO on Subaru (FOV = 2$^\prime \times 2^\prime$), and
4) MOIRCS on Subaru (FOV = 4$^\prime \times 7^\prime$).
Although the FOV of the above NIR cameras is much smaller than 
that of optical CCD cameras (e.g., $34^\prime \times 27^\prime$
for Suprime-Cam on Subaru),  
it seems important to promote a wide-field, narrowband imaging survey
to search for objects at $z > 7$.  MOIRCS (Tokoku et al. 2003) will be
one of most useful NIR cameras for coming years because of its largest FOV.
Another interesting new instrument is ^^ ^^ DAzLE (= The Dark Ages $z$ (redshift)
Lyman$\alpha$ Explorer", which is a special-purpose NIR narrowband
differential imager for VLT (Horton et al. 2004). 
Since this uses pairs of high-resolution ($R=1000$) narrowband filters closely
separated in wavelength, its limiting sensitivity is down to $\sim 2 \times
10^{-18}$ erg s$^{-1}$ cm$^{-2}$, corresponding to Ly$\alpha$ flux from 
a star-forming galaxy with SFR$ \sim$ a few $\times M_\odot$ y$^{-1}$ at
$z > 7$. Its FOV is also wide, $6.83^\prime \times 6.83^\prime$.

We hope that these new challenges from ground-based telescopes will be
able to probe real early phase of galaxy formation together with
Population III objects before going to JWST (The James Webb Space Telescope)
that will be a final 
weapon for our investigations of the early universe (Windhorst et al. 2005).


\label{lastpage}

\end{document}